\documentclass[twocolumn]{aastex631}
\usepackage{mathtools}
\usepackage{amsmath}
\usepackage{soul}
\usepackage{lipsum}
\usepackage{hyperref}

\newcommand\cms{$\mathrm{cm\ s^{-1}}$}
\newcommand\ms{$\mathrm{m\ s^{-1}}$}
\newcommand\kms{$\mathrm{km\ s^{-1}}$}

\newcommand\packagename{\texttt{SolAster}}
\newcommand\bobs{$|\hat{B}_{\mathrm{obs}}|$}
\newcommand\packageurl{https://tamarervin.github.io/SolAster/}
\newcommand\urlfootnote{\protect\footnote[1]{\url{\packageurl}}}
\newcommand\espressofoot{\protect\footnote[1]{\url{https://www.eso.org/sci/software/pipelines/espresso/espresso-pipe-recipes.html}}}

\shorttitle{Improving our understanding of stellar activity using space-based solar data}
\shortauthors{Ervin et al.}

\graphicspath{{./}{}}

\begin{document}

\title{Leveraging space-based data from the nearest Solar-type star to better understand stellar activity signatures in radial velocity data.}

\author[0000-0002-8475-8606]{Tamar Ervin}
\affiliation{Department of Physics and Astronomy, University of California, Los Angeles, Los Angeles, CA 90024, USA}
\affiliation{Jet Propulsion Laboratory, California Institute of Technology, 4800 Oak Grove Drive, Pasadena, CA 91109, USA
}

\author[0000-0003-1312-9391]{Samuel Halverson}
\affiliation{Jet Propulsion Laboratory, California Institute of Technology, 4800 Oak Grove Drive, Pasadena, CA 91109, USA
}

\author[0000-0002-5922-4469]{Abigail Burrows}
\affiliation{Department of Physics and Astronomy, Dartmouth College, Hanover, NH 03755, USA
}
\affiliation{Jet Propulsion Laboratory, California Institute of Technology, 4800 Oak Grove Drive, Pasadena, CA 91109, USA
}

\author{Neil Murphy}
\affiliation{Jet Propulsion Laboratory, California Institute of Technology, 4800 Oak Grove Drive, Pasadena, CA 91109, USA
}

\author[0000-0001-8127-5775]{Arpita Roy}
\affiliation{Space Telescope Science Institute, 3700 San Martin Dr., Baltimore, MD 21218, USA}
\affiliation{Department of Physics and Astronomy, Johns Hopkins University, 3400 N.\ Charles St., Baltimore, MD 21218, USA}

\author[0000-0001-9140-3574]{Raphaelle D.\ Haywood}
\affiliation{Astrophysics Group, University of Exeter, Exeter EX4 4QL, UK
}

\author[0000-0002-0594-7805]{Federica Rescigno}
\affiliation{Astrophysics Group, University of Exeter, Exeter EX4 4QL, UK
}

\author[0000-0003-4384-7220]{Chad F.\ Bender}
\affiliation{Steward Observatory, The University of Arizona, 933 N.\ Cherry Ave., Tucson, AZ 85721, USA}

\author[0000-0002-9082-6337]{Andrea S.J.\ Lin}
\affiliation{Department of Astronomy \& Astrophysics, 525 Davey Laboratory, The Pennsylvania State University, University Park, PA, 16802, USA}
\affiliation{Center for Exoplanets and Habitable Worlds, 525 Davey Laboratory, The Pennsylvania State University, University Park, PA, 16802, USA}

\author[0000-0002-0040-6815]{Jennifer Burt}
\affiliation{Jet Propulsion Laboratory, California Institute of Technology, 4800 Oak Grove Drive, Pasadena, CA 91109, USA
}

\author[0000-0001-9596-7983]{Suvrath Mahadevan}
\affiliation{Department of Astronomy \& Astrophysics, 525 Davey Laboratory, The Pennsylvania State University, University Park, PA, 16802, USA}
\affiliation{Center for Exoplanets and Habitable Worlds, 525 Davey Laboratory, The Pennsylvania State University, University Park, PA, 16802, USA}

\begin{abstract}

Stellar variability is a key obstacle in reaching the sensitivity required to recover Earth-like exoplanetary signals using the radial velocity (RV) detection method. To explore activity signatures in Sun-like stars, we present {\packagename}, a publicly-distributed analysis pipeline {\urlfootnote} that allows for comparison of space-based measurements with ground-based disk-integrated RVs. Using high spatial resolution Dopplergrams, magnetograms, and continuum filtergrams from the Helioseismic and Magnetic Imager (HMI) aboard the \emph{Solar Dynamics Observatory} (\emph{SDO}), we estimate \lq{}Sun-as-a-star\rq{} disk-integrated RVs due to rotationally modulated flux imbalances and convective blueshift suppression, as well as other observables such as unsigned magnetic flux. Comparing these measurements with ground-based RVs from the NEID instrument, which observes the Sun daily using an automated solar telescope, we find a strong relationship between magnetic activity indicators and RV variation, supporting efforts to examine unsigned magnetic flux as a proxy for stellar activity in slowly rotating stars. Detrending against measured unsigned magnetic flux allows us to improve the NEID RV measurements by $\sim$20\% ($\sim$50 {\cms} in a quadrature sum), yielding an RMS scatter of $\sim$60 {\cms} over five months. We also explore correlations between individual and averaged spectral line shapes in the NEID spectra and \emph{SDO}-derived magnetic activity indicators, motivating future studies of these observables. Finally, applying {\packagename} to archival planetary transits of Venus and Mercury, we demonstrate the ability to recover small amplitude ($<$50 {\cms}) RV variations in the \emph{SDO} data by directly measuring the Rossiter–McLaughlin (RM) signals.
\end{abstract}

\section{Introduction} \label{sec:intro}

The field of exoplanet science has drastically grown in popularity and fervor since the first confirmed exoplanet discovery around a Sun-like star \citep{Mayor-1995}. This has led to significant advancements in both instrumentation and data analysis techniques as we push towards the detection of a Earth-like planets \citep{Fischer-2016}. The radial velocity (RV) technique has been a cornerstone of exoplanet science since the first exoplanet discovery using Doppler velocimetry, and is credited with over 1000 additional planet discoveries \citep{Hatzes-2016, Fischer-2016}. 

The RV technique searches for periodic Doppler shifts in the host star’s spectra \citep{Hatzes-2016}. These periodic variations are driven by the presence of a planetary companion whose motion shifts the spectroscopic signature of its host star. As the field strives towards the detection of smaller, terrestrial-mass planets, improvements in RV measurement precision are required \citep{Fischer-2016} to push beyond the current $\sim$1 {\ms} measurement floor. Recent advancements in RV instrumentation, culminating in the delivery of a new generation of Doppler measurement facilities such as NEID \citep{Schwab-2016}, ESPRESSO \citep{Pepe-2021}, EXPRES \citep{Jurgenson-2016}, HARPS3 \citep{Thompson-2016}, and KPF \citep{Gibson-2018} aim to push down to the $\sim$30 {\cms} range. Detecting an Earth-like planet orbiting a Sun-like star requires additional improvement down to the 10 {\cms} level \citep{Wright-2018, Fischer-2016, Hatzes-2016}. 

The next challenge to improving detection sensitivity lies largely in the removal of stellar variability, which leads to noise that can often dominate measured RV variability \citep{Saar-1997} at the {\ms} level \citep{Crass-2021}. The signal from stellar activity can often mask or even masquerade as planetary signals \citep{Robertson-2015, Wright-2018}. Stellar activity signals are due to a combination of (super) granulation \citep{Meunier-2015, Dumusque-2011}, oscillations \citep{Palle-1995}, meridional circulation \citep{Meunier-2020}, magnetic activity, and photospheric motion \citep{Meunier-2010a, Haywood-2016, Crass-2021}. These phenomena are often periodic, aligning with the stellar rotation period and subsequent harmonics, and can consequently be mistaken for planetary signals \citep{Boisse-2011}. The lack of temporal stability across the stellar surface, coupled with inhomogeneous stellar intensity and differential rotation of the star, makes it difficult to robustly disentangle stellar signals from planetary ones when studying disk-integrated spectra.

To improve our understanding of stellar activity and its effects in Sun-like stars, we turn to our closest Solar-type star -- the Sun. The sheer amount of available solar data products, combined with established abilities to leverage high-cadence images of the solar surface to produce maps of solar velocity, intensity, and magnetic field strength, make the Sun the perfect candidate for studying activity-induced temporal variability \citep{SDO, HMI}. Using the Sun as a test case allows us to more cleanly separate the various component velocities and analyze relationships between measured disk-integrated RV variations and calculated solar observables. 

NASA’s Helioseismic and Magnetic Imager (HMI), an instrument aboard the \emph{Solar Dynamics Observatory} (\emph{SDO}), was built as the successor to the Michelson Doppler Imager (MDI) to study the solar surface magnetic field. Launched in 2010, it continuously observes the Sun in the spectral region of the Fe I 6173 {\AA} line, providing four high resolution data products: line-of-sight and vector magnetograms, continuum filtergrams, and Dopplergrams \citep{SDO, HMI}. These measurements of the magnetic field variability, intensity continuum, and velocity profile across the solar disk allow us to study the Sun's temporal variability and the effect these variations have on solar RV's \citep{SDO, HMI, Haywood-2016}. 

In this study, we develop a \emph{SDO}/HMI data analysis pipeline to compliment future extreme-precision RV (EPRV) studies of the Sun. Originally developed by \citet{Fligge-2000}, this technique has been adapted by \citet{Meunier-2010b}, \citet{Haywood-2016}, and \citet{Milbourne-2019} to extract disk-averaged quantities from spatially resolved solar observations. Our publicly available Python pipeline, \packagename, uses data products from \emph{SDO}/HMI to better characterize a suite of solar magnetic activity parameters, and performs a simple decorrelation analysis on disk-integrated solar RV measurements (now available from a number of RV facilities). There are two primary activity effects that strongly impact the measured RV (on timescales of days to months): the velocity variation due to the traversing motion of sunspots and faculae across the rotating solar surface, and the variation due to the suppression of the convective blueshift by active regions \citep{Aigrain-2012}. When linearly combined, these velocity components can be used to generate an independent estimate of the disk-integrated solar RV \citep{Haywood-2016}. The individual velocity components serve as a strong proxy for surface magnetic activity, providing a independent window into the stellar surface that can aid in interpreting ground-based RV measurements \citep{Haywood-2016, Milbourne-2019, Haywood-2020}. Additionally, from the \emph{SDO}/HMI data we calculate an array of magnetic observables that can be used to gauge the effects of the size and intensities of active regions on measured RV variations. 

The paper is organized as follows. In Section \ref{sec: sdo-hmi}, we describe our SDO/HMI data processing pipeline and analysis products. We outline the data correction process and methodology for classifying different magnetically active regions. In Section \ref{sec: rv-calc} we discuss the calculation of the full \lq{}Sun-as-a-star\rq{} RVs, outlining how each of the velocity components are independently calculated. We also describe the calculation of solar magnetic observables (Sections \ref{sec: Bobs} and \ref{sec: f}) and compare these results with our space-based measurements from HMI and ground-based RV measurements from the NEID instrument (\ref{sec: lines}). Finally, we apply these calculation techniques to archival planetary transits in Section \ref{sec: transits} to highlight the precision of the reconstructed RVs delivered by the pipeline, and demonstrate that magnetic variability can affect precision RV measurements at the 10's of {\cms} level over multi-hour timescales.

\section{{\packagename} - An \emph{SDO}/HMI analysis pipeline} \label{sec: sdo-hmi}

The plethora of data available from the Helioseismic and Magnetic Imager aboard \emph{SDO} allows us to calculate space-based, \lq{}Sun-as-a-star\rq{} radial velocity estimates that can be directly compared to ground-based measurements. Here we describe the underlying data products and techniques used to calculate various solar observables using {\packagename}.

Before computing the RVs from the \emph{SDO}/HMI data, there are a number of data preparation steps required. The suite of \emph{SDO}/HMI images used in this study were: wide-band continuum filtergrams (intensity grams), line-of-sight longitudinal magnetic field measurements (magnetograms), and maps of solar surface velocity (Dopplergrams) \citep{SDO, HMI} (see Figure~\ref{fig: sdo/hmi} for example images). These three data products provide the necessary intensity, magnetic field strength, and velocity information for active regions to be detected, tracked, and accurately integrated into the full RV model. 

HMI data products are publicly available and can be queried from the data archive using \texttt{Sunpy}, a community based Python package for solar data analysis \citep{Sunpy}. In addition to providing archive querying capabilities, {\texttt{Sunpy}} includes user-friendly methods for accessing and visualizing solar data. 

Using {\texttt{Sunpy}, {\packagename} calculates a combination of velocities and magnetic observables using the \emph{SDO}/HMI intensity, velocity, and magnetic field data. Photometric and convective velocity components are independently calculated and then linearly combined to generate \lq{}Sun-as-a-star\rq{} RVs. Additionally, we calculate both unsigned magnetic flux and filling factor, which can be used to study the correlation between disk-integrated radial velocity and measures of magnetic activity. These magnetic observables are calculated contemporaneously to the RV calculations and include unsigned flux and filling factor measurements specific to all relevant active regions (plage, intranetwork, and sunspots). We also look at flux due to convective regions, and area cuts to study the differing effects between large and small active regions on the modeled RVs.

\subsection{Coordinate transformations}

Before calculating the three-dimensional Heliocentric velocity, the data must be transformed from the Helioprojective Cartesian Frame into the Heliographic Carrington frame, then corrected for line-of-sight projections and relative positioning of the spacecraft. This transformation is based off the description in \citet{Thompson-2006} and is necessary to ensure the images are centered on the solar surface and independent of the Carrington rotation cycle, the 25.38 day solar sidereal rotation period \citep{Carrington-1859}. To transform coordinate systems, we rotate the image grid from Cartesian pixel coordinates to Heliographic Carrington coordinates by building a rotation matrix calculated from the reference coordinates listed in each image's \texttt{FITS} header \citep{Ulrich-2006}. Each HMI image's relative pixel locations in the Heliographic Carrington frame are specified by $(w_{ij},\, n_{ij},\, r_{ij})$ denoting the direction westward, northward, and radially outward from disk center, respectively. This coordinate system fixes the image onto the solar surface and allows for a determination of the relative position of the spacecraft with respect to the Sun using only the radial coordinate. Additionally, we constructed an array of $\mu \; (\cos\theta)$ values for each pixel in each image, which determines the position of the pixel relative to disk center. Flux values for pixels with $\mu$ values below 0.3 were set to zero in all images as the limb-brightening model is often unreliable far from disk center \citep{Haywood-2016}, and projection issues can cause non-physical fluctuations in measured values. 

\subsection{Spacecraft Velocity Correction} \label{sec: vsc}

To isolate the solar velocity component due to magnetic activity alone, we first corrected the Dopplergrams for the relative motion of the spacecraft.

We corrected the Dopplergrams for the motion of the spacecraft relative to the Sun by building a pixel-wise mask of the relative spacecraft velocity. The w, n, and r components of the relative spacecraft velocity are read in from the \texttt{FITS} headers with a quoted precision of $0.01$ {\ms} \citep{Hoeksema-2018}. We then calculated the position of the spacecraft relative to each pixel $ij$, which combined with the velocity components from the \texttt{FITS} header, determined the required velocity correction. After the coordinate transformation, the spacecraft is located at position $(0, \, 0, \, r_{\mathrm{sc}})$ where $r_{\mathrm{sc}}$ is the radial position of the spacecraft relative to disk-center, and can be determined by dividing the net distance to the Sun by the solar radius (both these values are found in the \texttt{FITS} header of all \emph{SDO}/HMI images with keywords \texttt{dsun\_obs} and \texttt{rsun\_ref}).

\begin{equation}
    \begin{split}
        \delta w_{ij} = w_{ij} - 0 \\
    \delta n_{ij} = n_{ij} - 0 \\
    \delta r_{ij} = r_{ij} - r_{\mathrm{sc}}
    \end{split} 
\end{equation}

The $(w, \, n, \, r)$ components due to the relative motion of the spacecraft are found in the \texttt{FITS} header of the Dopplergram files (\texttt{obs\_vw}, \texttt{obs\_vn}, \texttt{obs\_vr} respectively). We then project these components such that each pixel $ij$ has a Heliocentric velocity magnitude of \citep{Haywood-2016}:
	
\begin{equation}
\begin{multlined}
    v_{\mathrm{sc}, \, ij} = - \frac{\delta w_{ij} \, v_{\mathrm{sc}, \, w_{ij}} + \delta n_{ij} \, v_{\mathrm{sc}, \, n_{ij}} + \delta r_{ij} \, v_{\mathrm{sc}, \, r_{ij}}}{d_{ij}}
\end{multlined}
\end{equation} where $d_{ij}$ is the distance between the spacecraft and pixel $ij$.

\begin{equation}
    d_{ij} = \sqrt{\delta w^2_{ij} + \delta n^2_{ij} + \delta r^2_{ij}}
\end{equation}

\subsection{Solar Rotational Velocity Correction} \label{sec: vrot}

Next, we turn our attention to the differential rotation of the solar disk which must be accounted for when correcting the measured Doppler maps. Differential rotation is the result of turbulent motion and convective activity due to temperature gradients permeating outwards from the stellar core \citep{Schou-1998}. This produces a latitude-based rotation profile, where the rate of surface rotation is maximized at the equator ($\phi = 0 ^{\circ}$) and is inversely proportional to latitude \citep{Schroter-1985}. The angular velocity due to rotation in the photospheric layer ranges from 14.1-14.4 deg day$^{-1}$ at the equator to 10.07 deg day$^{-1}$ at the poles \citep{Snodgrass-1984}. The sidereal rotation period for a Carrington rotation is 25.38 days, which is the rotation rate at a latitude of 26$^{\circ}$, where sunspots are most often found. This rotation period is accounted for in our coordinate transformation to the Heliographic Carrington frame \citep{Thompson-2006}.  

\citet{Snodgrass-1990} used full-disk Magnetograms and Dopplergrams from the Mount Wilson Observatory to track magnetic features on the solar surface over time in order to build a model of the solar differential rotation profile. They determined three constants $\alpha_1, \, \alpha_2, \, \alpha_3$, all of which are in units of deg day$^{-1}$. The parameterization of the differential rotational profile is as follows:

\begin{equation}
    \omega(\phi) = \alpha_1  - \alpha_2 \, \sin^2(\phi) - \alpha_3 \, \sin^4 (\phi).
\end{equation} 

Using this parameterization with coefficients 14.713, 2.293, and 1.787 as $\alpha_1, \alpha_2, \alpha_3$ respectively, we calculate our differential rotation profile and project this onto the solar disk to build a map of solar rotational velocity at each latitude. We then project this differential rotation profile into the Heliographic Carrington frame to determine the rotational velocity component at each pixel. Finally, we calculate the full rotational velocity array based on the methods of \citet{Haywood-2016} and \citet{Milbourne-2019}: 

\begin{equation}
    v_{\mathrm{rot}} = - \frac{\delta w_{ij} \, v_{\mathrm{rot}, \, w_{ij}} + \delta n_{ij} \, v_{\mathrm{rot}, \, n_{ij}} + \delta r_{ij} \, v_{\mathrm{rot}, \, r_{ij}}}{d_{ij}}.
\end{equation} 

\subsection{Foreshortening Correction} \label{sec: unsigned-field}

The line-of-sight magnetograms measure the longitudinal surface magnetic field. Foreshortening causes a decrease in observed spatial resolution relative to the distance from disk center due to the geometric projection and must be accounted for when estimating the true magnetic flux \citep{Zhao-2016}. This measured magnetic field is less than the true radial solar magnetic field by a factor of $\mu = \cos{(\theta)}$, where $\theta$ is the center-to-limb angle \citep{Zhao-2016}. To calculate the true field strength, we divide the observed field ($B_\mathrm{obs}$) by $\mu$ and recover the full radial field:

\begin{equation}
    B_{r, \, ij} = B_{\mathrm{obs}, \, ij} / \mu_{ij}.
\end{equation}

Additionally, we set all pixels with magnetic field strengths below the noise threshold ($\sigma_{B_{\mathrm{obs}, \, ij}}$) to zero to account for instrument noise as described in \citet{Yeo-2013}. Although the noise does increase as a function of angle from disk center ($\mu$) we take a constant minimum noise threshold of 8G based on \citet{Yeo-2013}. Therefore, pixels with longitudinal magnetic field strengths ($B_{\mathrm{obs}, \, ij}$) below 8G are set to 0 for both $B_{\mathrm{obs}, \, ij}$ and $B_{r, \, ij}$. This ensures our magnetic measurements are not contaminated by instrument noise, which would otherwise propagate through many aspects of the analysis pipeline. 

\subsection{Limb Darkening Correction}\label{sec: lbc}

Similar to the effect foreshortening has on the HMI magnetograms, the continuum images are also affected by limb-darkening. We correct for this by using a static fifth-order circularly symmetric polynomial brightness function ($L_{ij}$), with scaling coefficients determined through empirical methods by \citet{Allen-1973}. This polynomial produces a pixel-wise array of correction values and the base intensity image is divided by these correction factors. 
\begin{equation}
    I_{\mathrm{flat}, \, ij} = \frac{I_{ij}} {L_{ij}}
\end{equation}
The flattened intensity image can now be used to classify bright and dark regions (Figure \ref{fig: sdo/hmi}), which are separated via thresholding. 

\subsection{Region Identification} \label{sec: regions}

The underlying assumption in our space-based RV calculation is that magnetic activity is the primary driver of bulk RV variability in the Sun. We identify magnetically active regions, and differentiate between regions of bright faculae and dark sunspots, to distinguish the impact of different types of magnetic activity on RVs.

Active regions are detected using a thresholding identification scheme described in \citet{Yeo-2013}. Regions above the threshold are marked as \lq{}active\rq{} and regions below the threshold are stored as \lq{}quiet-Sun\rq{} pixels. Additionally, we remove pixels near the solar limb ($\mu < 0.1$) and ignore pixels with $\mu$ values below 0.3 since the limb-darkening model is often flawed near the limb, as was done in \citet{Haywood-2016} and \citet{Milbourne-2019}. We apply the same magnetic threshold described in \citet{Yeo-2013}, where pixels three times the noise cutoff in unsigned radial magnetic field strength (8G) are considered active:
\begin{equation}
    |B_{r, \, \mathrm{thresh}}| = 3 \sigma_{B_{\mathrm{obs}, \, ij}} / \mu_{ij}
\end{equation} where $\sigma_{B_{\mathrm{obs}, ij}}$ is the magnetic noise level of 8G from \citet{Yeo-2013}. We set any isolated active pixels, i.e. those with no identified neighboring active pixels, to 0 (\lq{}quiet-Sun\rq{}) as these can often be misidentified as sunspots and may instead be instrumental artifacts. 

Once active regions are identified (Fig \ref{fig: sdo/hmi}), we then apply intensity thresholding to differentiate between faculae and sunspot regions. Similar to the magnetic thresholding previously described, we base our intensity thresholding on values determined by \citet{Yeo-2013} and used by \citet{Haywood-2016} and \citet{Milbourne-2019}. Pixels with flattened intensity values above the threshold are denoted as faculae and those below the threshold are sunspots: 

\begin{equation}
    I_\mathrm{thresh} = 0.89 \; I_\mathrm{quiet}.
\end{equation}

The intensity threshold is based on $I_\mathrm{quiet}$, the mean flattened pixel intensity of quiet-Sun pixels, and is calculated by summing the flattened intensity of quiet-Sun pixels with a binary weighting array based on magnetic thresholding:

\begin{equation}
    I_\mathrm{quiet} = \frac{\sum_{ij} I_{\mathrm{flat}, \, ij} \, W_{ij}}{\sum_{ij} W_{ij}},
\end{equation}

where $W_{ij}$ is set to 1 for quiet-Sun pixels ($|B_{r, \,ij}| < |B_{r, \, \mathrm{thresh}, \, ij}|$) and 0 for active pixels.

\begin{figure*} [htb!]
    \includegraphics[width=\textwidth]{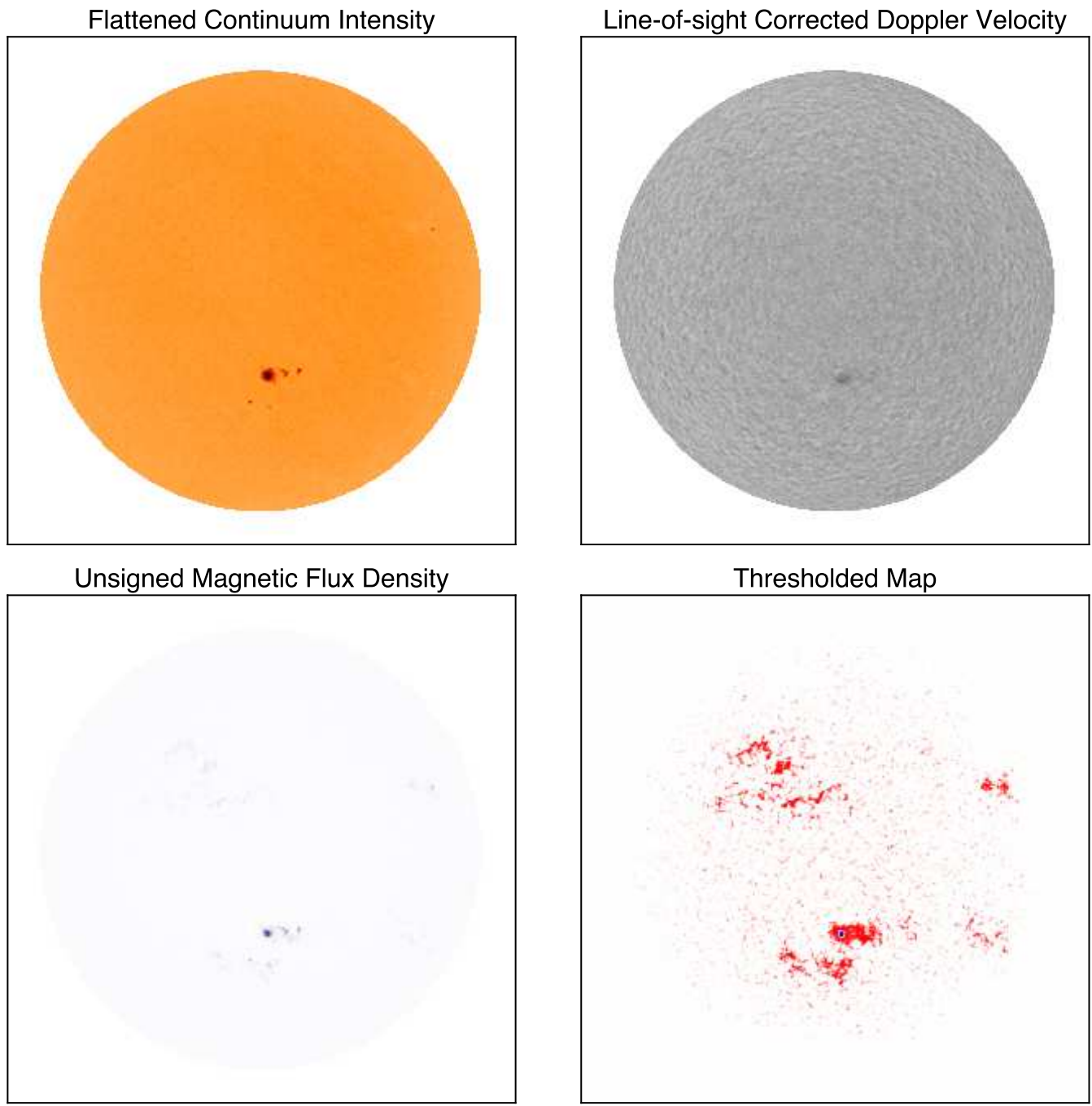}
    \caption{Example of corrected \emph{SDO}/HMI data products during the highest activity day (largest magnetic filling factor) in our studied time frame (June 30, 2021). \emph{Top left:} Flattened continuum intensity after limb-darkening correction (used to identify bright and dark regions) \emph{Top right:} Doppler velocity map after removal of spacecraft and solar rotational velocity. \emph{Bottom left:} Unsigned magnetic flux density ($|B_{r}|$). In this image, 3.25\% of the Sun is covered by faculae/plage and 0.026\% by sunspots, as identified by our pipeline. \emph{Bottom right:} Thresholded image showing magnetically active regions identified by our algorithms. Faculae/plage regions are red and sunspot regions are blue.}
    \label{fig: sdo/hmi}
\end{figure*}

\subsection{Radial Velocity Calculation} \label{sec: rv-calc}

Following \cite{Haywood-2016}}, we parameterize the full, disk-integrated solar radial velocity as a linear combination of contributions from the quiet-Sun and active regions. Active regions produce RV variations through two primary mechanisms: photometric effect and convective effect. Meanwhile, the quiet-Sun RVs are primarily driven by granulation. 

\subsection{Photometric Contribution} \label{sec: vphot}

The photometric velocity traces the rotational Doppler imbalance caused by bright faculae and dark sunspots. The presence of bright and dark active regions leads to an inhomogeneity across the solar disk, altering the Doppler balance between redshifted and blueshifted hemispheres. This leads to RV shifts of up to several percent depending on spot size and stellar activity levels \citep{Saar-1997}. Sunspots are generally the dominant source of variability in the photometric RV signal, although plage regions also contribute to the signal \citep{Lagrange-2010}. The photometric effect due to these two factors is accounted for in$\Delta \hat{v}_\mathrm{phot}$, which we calculate based on the methodology outlined in \citet{Haywood-2016} and \citet{Milbourne-2019}: 
\begin{equation}
	\Delta \hat{v}_\mathrm{phot} = \frac{\sum_{ij} v_{\mathrm{rot}, \, ij}\, (I_{ij} - \hat{K}\,L_{ij})\, W_{ij}}{\sum_{ij} I_{ij}},
\end{equation} where $\hat{K}$ is a scaling factor based on the limb darkening correction polynomial:
\begin{equation}
	\hat{K} = \frac{\sum_{ij} I_{ij} \, L_{ij}\, W_{ij}}{\sum_{ij} L^2_{ij} \, W_{ij}}.
\end{equation}

$I_{ij} - \hat{K}\,L_{ij}$  is the intensity map corrected for limb-darkening, as seen in the top left panel of Figure~\ref{fig: sdo/hmi}. An example of the $W_{ij}$ weighting array can be seen in the bottom right panel of Figure~\ref{fig: sdo/hmi}, where $W_{ij} = 0$ for quiet-Sun pixels and $W_{ij} = 1$ for active pixels.

The velocity perturbations from faculae and sunspots are approximately anti-correlated due to their opposing flux signs. When calculating the photometric velocity component, we find that this velocity perturbation is almost entirely driven by sunspots (see Figure~\ref{fig: magnetic}), corroborating the results of \citet{Meunier-2010a}. This is likely due to the Sun’s geometric configuration and the ratio of bright/spot regions at the time, meaning this may not be the case for other stars with different filling factors and distributions of bright and dark regions. 

\subsection{Convective Contribution} \label{sec: vconv}

Active magnetic regions have different velocity amplitudes and surface areas distributed between upward and downward flows of solar granulation \citep{Dravins-1990}. In the photosphere, active magnetic regions inhibit granular convective motions of the quiet-Sun, and these convective motions manifest as wavelength shifts of photospheric lines \citep{Dravins-1981}. While the photometric velocity variation is driven by sunspots, the convective velocity variation is driven by larger brighter faculae regions and thus these drive the overall RV signal. 

\emph{SDO}/HMI images can resolve these granules, allowing us to calculate the velocity contribution specifically due to suppression of the convective blueshift. In convective cells, dark outward flowing plasma at the cell’s center and downward flowing bright plasma on the cells edge leads to overall convective blueshifts on the order of 0.5 {\kms} \citep{Dravins-1981}. This correlation weakens across the solar disk as we see primarily horizontal velocity flows on the solar limb \citep{Dravins-1990}.

The effect of the suppression of the convective blueshift varies with line depth \citep{Gray-2009} and thus we expect to see different temporal convective velocity shifts across different wavelengths. For this reason, we do not expect perfect correlation between NEID observations and the space-based convective velocity, as the ground-based RV measurements utilize thousands of spectral features  while \emph{SDO}/HMI observes velocities only in the magnetically sensitive 6173.3 {\AA} Fe I line. We use linear regression to scale the \emph{SDO}/HMI derived convective velocity to account for this difference (see Section~\ref{sec: rv}). 

The convective velocity is then calculated by taking the disk averaged Doppler velocity, $\hat{v}$, and subtracting from it the disk averaged quiet-Sun velocity, $\hat{v}_\mathrm{quiet}$. We subtract the quiet-Sun velocity because \emph{SDO}/HMI Dopplergrams are not well calibrated nor stable over long timescales \citep{Haywood-2020}.
\begin{equation}
    \Delta \hat{v}_\mathrm{conv} = \hat{v} - \hat{v}_\mathrm{quiet}.
\end{equation}
We calculate the disk-averaged and quiet-Sun velocities following the methodology of \citet{Haywood-2016} and \citet{Milbourne-2019}:
\begin{equation}
    \hat{v} = \frac{\sum_{ij} (v_{ij} - v_{sc, \, ij} - v_{rot, \, ij}) \, I_{ij}}{\sum_{ij} I_{ij}}
\end{equation} where $v_{sc, \, ij}$ is the relative spacecraft velocity and $v_{rot, \, ij}$ is the solar rotational velocity. An example of the corrected spacecraft velocity ($v_{ij} - v_{sc, \, ij} - v_{rot, \, ij}$) can be seen in the top right panel of Figure~\ref{fig: sdo/hmi}. The quiet-Sun velocity is calculated by using the corrected Doppler velocity and weighting by the intensity of quiet-Sun pixels:

\begin{equation}
    \hat{v}_\mathrm{quiet} = \frac{\sum_{ij} (v_{ij} - v_{sc, \, ij} - v_{rot, \, ij}) \, I_{ij} \, W_{ij}}{\sum_{ij} I_{ij} \, W_{ij}}
\end{equation}where $W_{ij}$ is the magnetic weighting array.

\subsection{RV Reconstruction from Velocity Features} \label{sec: rv}

To estimate the full disk-integrated \emph{SDO}/HMI space-based radial velocities, we follow the methodology outlined in  \citet{Milbourne-2019} and \citet{Haywood-2020}, adapted from \citet{Haywood-2016}. We build a model radial velocity variation  $\Delta RV_\mathrm{model}$ assuming a linear combination of $\Delta \hat{v}_\mathrm{conv}$ and $\Delta \hat{v}_\mathrm{phot}$:
\begin{equation} \label{eq: rv_model}
    \Delta RV_\mathrm{model} = A\, \Delta \hat{v}_\mathrm{phot} +  B\, \Delta \hat{v}_\mathrm{conv} + RV_0
\end{equation}where $A$ and $B$ are independent scaling factors, and $RV_0$ is the relative RV offset parameter. Similar to \citet{Milbourne-2019}, these coefficients are determined by linear least-squares optimization using the ground-based RV measurements, assuming the two RV components are orthogonal and do not have correlated noise. These scaling factors account for the systematic differences between observations taken using \emph{SDO}/HMI in one line ($\mathrm \lambda = 6173.3 \, \mathrm {\AA}$), and ground-based spectra using thousands of lines.

\subsection{Validation of {\packagename} using previously published \emph{SDO}/HMI measurements} \label{sec: ref-rvs}
 
\begin{figure} [htb!]
    \includegraphics[width=\columnwidth]{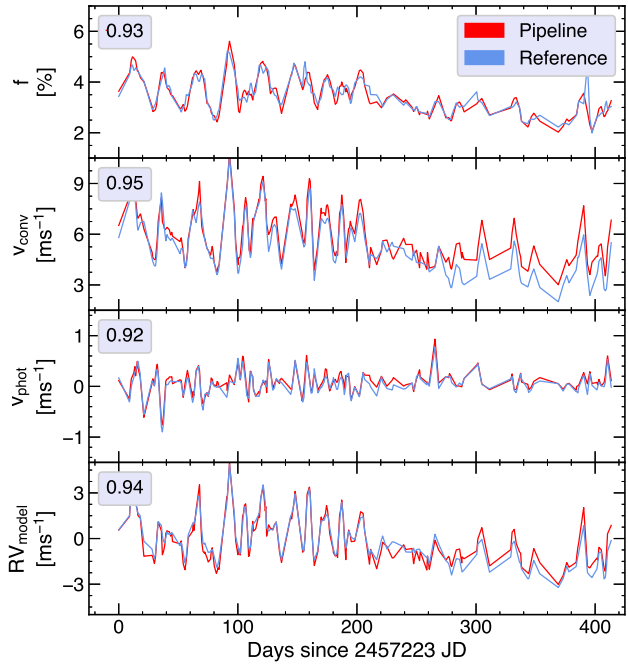}
    \caption{Comparison between our pipeline-derived \emph{SDO}/HMI derived velocities and observables (red) with reference calculations from \citet{Milbourne-2019} (blue). Correlation coefficients are shown in the upper left of each plot. \emph{First panel:} Filling factor estimate, defined as the percentage of magnetically active pixels on the solar surface. The filling factor, $f$, for this period reflects the period of higher solar activity during which this data was taken. \emph{Second/Third panel:} Comparison of unscaled velocity components $\Delta \hat{v}_\mathrm{conv}$ and $\Delta \hat{v}_\mathrm{phot}$. \emph{Fourth panel:} Total $\Delta RV_\mathrm{model}$ from \citet{Milbourne-2019} compared with our results, showing excellent agreement.}
    \label{fig: vel-mil-comp}
\end{figure}

As our methodology for calculating the velocity components and full model RVs is based on the methods outlined in \citet{Haywood-2016} and \citet{Milbourne-2019}, we analyzed the same solar data used in these studies to verify our performance. We use our pipeline to calculate RVs for the time frame in \citet{Milbourne-2019} to compare our model with published reference values. When comparing our derived \emph{SDO}/HMI measurements with the equivalent measurements in \citet{Milbourne-2019} we find excellent agreement. The Spearman correlation coefficients for $\Delta \hat{v}_\mathrm{conv}$, $\Delta \hat{v}_\mathrm{phot}$, and $\Delta RV_\mathrm{model}$ are 0.97, 0.93, and 0.97 accordingly (Figure \ref{fig: vel-mil-comp}). Our model RVs and unsigned flux both show strong correlation with the HARPS-N RV measurements with Spearman correlation coefficients of 0.80 and 0.74 respectively. These strong correlations show the reproducibility of the \citet{Haywood-2016} and \citet{Milbourne-2019} results using our analysis pipeline.

\section{Comparison of \emph{SDO}/HMI derived observables with ground-based measurements} \label{sec: comp}

Our ultimate goal in calculating these solar observables is to gain insight into the physical mechanisms driving measured RV variability seen in ground-based Doppler measurements. We use data from the recently-commissioned NEID instrument \citep{Schwab-2016}, which has a dedicated solar feed that delivers disk-integrated sunlight to the RV spectrometer (Lin et al. 2021 submitted.). NEID records high signal-to-noise (SNR$\sim$600) spectra every $\sim$90 seconds throughout the day. NEID spectra are reduced using the standard NEID pipeline, which delivers both integrated RVs and cross-correlation functions (CCFs) for each frame recorded throughout the day \footnote{\url{https://neid.ipac.caltech.edu/search_solar.php}}. We filtered for days with low cloud coverage, using data from the  pyheliometer atop the NEID Solar telescope (Lin et al. 2021 submitted.), and good instrumental drift correction. 

Using our \emph{SDO} analysis pipeline, we then computed the component RVs ($\Delta \hat{v}_\mathrm{phot}$ and $\Delta \hat{v}_\mathrm{conv}$) during periods when NEID spectra were being collected. The independent amplitudes of $\Delta \hat{v}_\mathrm{phot}$ and $\Delta \hat{v}_\mathrm{conv}$, are significantly lower for the period of NEID data collection (December 2020 - May 2021) than the values shown in \citet{Haywood-2016} and \citet{Milbourne-2019}, reflecting the low level of magnetic activity over the period for which NEID has been observing the Sun (Table~\ref{tab: basis}).
 
We find that the resultant \lq{}Sun-as-a-star\rq{} \emph{SDO}/HMI computed RVs are largely dominated by the convective velocity signal, likely due to the low level of surface features (spots, plages, etc.) on the solar surface during the period of analysis. This is further highlighted when comparing the modeled RVs and convective velocity component with unsigned flux and filling factor, which are all strongly correlated (Figure~\ref{fig: vel}).

\begin{table*} 
    \centering 
    \resizebox{.8\textwidth}{!}{%
    \begin{tabular}{c c c c} 
        \hline
        Velocity component & Velocity RMS & \citet{Haywood-2016} & \citet{Milbourne-2019} \\ [0.5ex] 
        \hline\hline
        $\Delta \hat{v}_\mathrm{phot}$ & .06 m s$^{-1}$ & $0.17$ m s$^{-1}$& $0.21$ m s$^{-1}$\\ 
        $\Delta \hat{v}_\mathrm{conv}$ & 0.62 m s$^{-1}$ & $1.30$ m s$^{-1}$ & $0.88$ m s$^{-1}$\\ 
        Ground-based RVs & 0.8 m s$^{-1}$ (NEID) & 3.12 {\ms} (HARPS) & 1.64 {\ms} (HARPS-N) \\
        \hline
    \end{tabular}
    }
    \caption{RMS amplitudes for full time series of NEID data collection, (December 2020 - May 2021) compared with archival results from \citet{Haywood-2016} and \citet{Milbourne-2019}. The $\Delta \hat{v}_\mathrm{phot}$ and  $\Delta \hat{v}_\mathrm{conv}$ values are derived from \emph{SDO}/HMI products. The lower levels of solar activity during the period of NEID data collection are readily apparent when comparing to the HARPS-N data from \citet{Haywood-2016} and \citet{Milbourne-2019}, which covered a more active span of the Sun. }
    \label{tab: basis}
\end{table*}

\subsection{Comparison of SDO Model RVs with NEID Ground Based RVs} \label{sec: neid-rvs}

We then proceeded to compare the results of the \emph{SDO}/HMI measurements to NEID ground-based RVs collected during instrument commissioning between December 2020 and May 2021 (the nominal NEID commissioning period). Using these data, we refit for the linear coefficients in equation~\ref{eq: rv_model} for the convective and photometric velocity components. Table~\ref{tab: scale} lists our derived scaling factors using the NEID RVs for calibration in comparison to the scaling factors derived in \citet{Haywood-2016} and \citet{Milbourne-2019}. Both scaling factors characterize the impact of a measurement taken using one line (\emph{SDO}/HMI data) to ground-based RVs measured across many lines, and we expect these factors to change over time due to fluctuations in spot coverage and activity level \citep{Milbourne-2019}. Scaling factor A traces the contribution of rotating active regions (primarily spots) to the bulk RVs. Our time frame of interest had consistently low spot coverage, leading to a very low amplitude for the photometric component, and thus large uncertainty in the determination of scaling factor A (see Table~\ref{tab: scale}). Scaling factor B is a measure of the systematic difference in the convective blueshift due to varying spectral line formation depths.   

\begin{figure} [htb!]
    \includegraphics[width=\columnwidth]{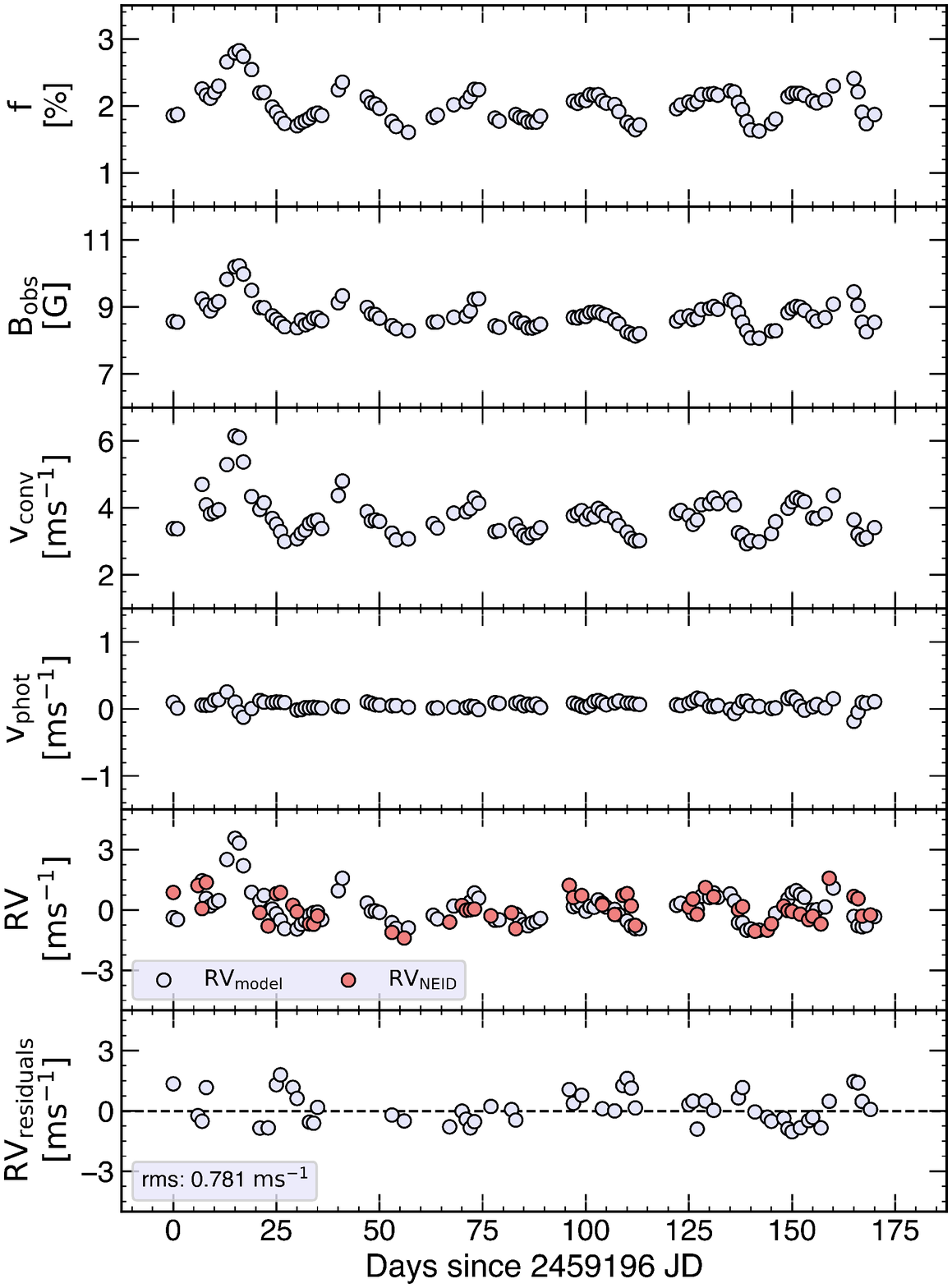}
    \caption{Comparison between \emph{SDO}/HMI derived observables and NEID RV observations. \emph{Top panel:} Filling factor computed from \emph{SDO}/HMI magnetograms. \emph{Second panel:} Unsigned magnetic flux computed from \emph{SDO}/HMI magnetograms. \emph{Third/Fourth panel:} Unscaled basis velocity components $\Delta \hat{v}_\mathrm{conv}$ and $\Delta \hat{v}_\mathrm{phot}$. \emph{Fifth panel:} Total $\Delta RV_\mathrm{model}$ computed for the NEID observational time span (in light lavender) with the ground-based NEID RVs filtered for bad weather and cloud cover overlaid (in dark red). The average photon-limited error on the daily binned $ RV_\mathrm{NEID}$ measurements is $\sim$5 {\cms}.  \emph{Sixth panel:} Residuals between $\Delta RV_\mathrm{model}$ and $\Delta RV_\mathrm{NEID}$ from panel five. The residual rms scatter is 78 {\cms}.}
    \label{fig: vel}
\end{figure}

As established by \citet{Meunier-2010a}, \citet{Haywood-2016}, and \citet{Milbourne-2019} we expect the suppression of the convective blueshift to dominate the overall RV, which is consistent with our measurements during this time period (see Table~\ref{tab: basis}). Our ground-based RVs from NEID do not show measurable correlation with the photometric velocity signals calculated from the \emph{SDO}/HMI images (Figure~\ref{fig: vel}), which is consistent with the current phase of solar activity (minimum) and in agreement with \citet{Lagrange-2010}. The lack of sunspots during the NEID commissioning period leads to the low variability in the photometric velocity, which further minimizes the contribution of the photometric component to the overall model RVs ($\Delta RV_\mathrm{model}$), as seen by the low value of scaling factor A in Table~\ref{tab: basis}. 

\begin{table*} 
    \centering
    \resizebox{.8\textwidth}{!}{%
    \begin{tabular}{c c c c}
        \hline
        Parameter & This work & \citet{Haywood-2016} & \citet{Milbourne-2019} \\
        \hline\hline
        A (photometric) & $1.10 \pm 1.07$ & $2.45 \pm 2.02$ & $2.24 \pm 0.60$  \\
        B (convective) & $1.42 \pm 0.11$ & $1.85 \pm 0.27$ & $0.93 \pm 0.11$  \\
        RV$_0$ ($\gamma$) & $-646.08 \pm 0.38$ m s$^{-1}$ & $99.80 \pm 0.28$ m s$^{-1}$ & $102.51 \pm 0.06$ m s$^{-1}$ \\
        \hline
    \end{tabular}
    }
    \caption{Scaling Factors for \emph{SDO}/HMI derived RVs using NEID data compared with previous results from \citet{Haywood-2016} and \citet{Milbourne-2019} using HARPS-N. In the period studied here, the overall activity level and amplitude of the velocity components was lower along with the rms amplitude of the RVs (see Table~\ref{tab: basis}). We see a $\sim$20\% difference between the $A$ and $B$ parameter values between the different time periods. This discrepancy is likely due to the significantly lower level of activity in our recent data from 2020-2021.} 
    \label{tab: scale}
\end{table*}

We are able to improve our results when restricting our analysis to days with the most reliable ground-based observations. These dates have both excellent observing weather (very low or no cloud coverage) and verified wavelength solutions from the NEID laser frequency comb. For these dates we see very strong correlation between our ground based RV measurements and both the model RVs and unsigned magnetic flux. We find that unsigned magnetic flux is a strong proxy for RV variation supporting the conclusion of \citet{Haywood-2020}.

\subsection{Active Region Area Dependence} \label{sec: area-active}

In addition to the comparison of $\Delta RV_\mathrm{model}$ with the convective velocity and ground-based measurements, we examine the area dependence of active regions on the suppression of the convective blueshift. To differentiate between large and small convective regions, we build an area plot as a function of latitude, similar to \citet{Milbourne-2019} and use 20 $\mu$Hem as our area threshold. We find network regions below the cutoff across the solar disk, while larger plage and spot regions are found near the solar equator at latitudes $\pm 30 ^\circ$. \citet{Milbourne-2019} implemented the same area cutoff and found similar latitude cuts of large active regions, $0.75 < \sin \Phi \leq 1.0$, where $\Phi$ is active region co-latitude.

\begin{figure} [htb!]
    \centering
    \includegraphics[width=\columnwidth]{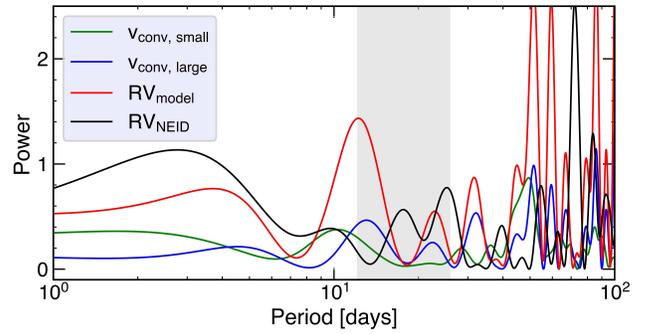}
    \caption{Periodogram showing the contributions due to the small active regions (green) and large active regions (blue) in comparison with model (red) and ground-based RV measurements (black). The small active regions do not contribute to power on rotation timescales, supporting results of \citet{Milbourne-2019}, while large active regions (spots/plage) show measurable power at periodicities near $P_\mathrm{rot}$.}
    \label{fig: area-vconv}
\end{figure}

We then compute the $\Delta \hat{v}_\mathrm{conv}$ for the full time series using the contributions from the small network regions, and large plage/spot regions separately to compare with the $RV_\mathrm{NEID}$  from NEID. In support of \citet{Milbourne-2019} we find that large plage and spot regions drive the RV variability more so than smaller network regions (Figure~\ref{fig: magnetic}). Additionally, we find that the power spectral density (PSD) contributions due to small active regions do not contribute on timescales relevant to this study (rotation timescales) as seen in Figure~\ref{fig: area-vconv}, similar to the results found in \citet{Milbourne-2019}, while large active regions (plage/spots) show rotational modulation as seen in Figure~\ref{fig: area-vconv}. While the convective velocity component is calculated with the corrected Doppler map (rotational velocity removed) there is still periodic structure on rotation timescales (see Figures~\ref{fig: vel-mil-comp} \& \ref{fig: vel}). Large active magnetic regions (plage, spots) are expected to drive variability on shorter timescales such as the period in this study, while the network is expected to show more impact on longer timescales such as a Solar Cycle \citep{Milbourne-2019}.


\subsection{Unsigned Magnetic Flux} \label{sec: Bobs}

Unsigned magnetic flux, {\bobs}, has been shown to be a valuable proxy for stellar activity in the Sun \citep{Haywood-2020}. Our pipeline calculates unsigned magnetic flux using the methodology posed in \citet{Haywood-2016} and \citet{Haywood-2020}, which yields a high signal-to-noise, independent activity metric to guide our analysis of ground-based RVs. Figure \ref{fig: vel} shows the calculated {\bobs} time series during the full NEID commissioning period. {\bobs} is determined by performing an intensity weighted sum of each pixel in the magnetogram:

\begin{equation}
	|\hat{B}_{\mathrm{obs}}| = \frac{\sum_{ij} |B_{\mathrm{obs}, \, ij}| I_{ij}}{\sum_{ij} I_{ij}}
\end{equation}

For our five month span (December 2021 - May 2021) of \lq{}best\rq{} weather dates, defined as days with no measurable cloud cover and minimal extinction at Kitt Peak, we find a Spearman correlation coefficient of 0.43 between {\bobs} and measured NEID RVs and a correlation of 0.29 between measured NEID RVs and model RVs. We see strong correlation between {\bobs} and both the space-based RVs and $\Delta \hat{v}_\mathrm{conv}$ ($\sim$ 0.90) throughout the entire time span, emphasizing the dependence of magnetic activity on the suppression of the convective blueshift. This is expected based on previous works such as \citet{Meunier-2010a}, \citet{Meunier-2010b}, \citet{Haywood-2016}, and \citet{Haywood-2020}. 

These observations were taken during the solar minima of Solar Cycle 25 and thus there was very little to no sunspot activity on the solar surface. The convective velocity is primarily driven by large magnetic structures, specifically plage regions. We see strong correlations between the plage filling factor, convective velocity, and unsigned flux. Additionally, we find the unsigned flux due to active regions strongly correlates with the \emph{SDO}/HMI model RVs. 

\begin{figure*} [htb!]
    \centering
    \includegraphics[width=0.9\textwidth]{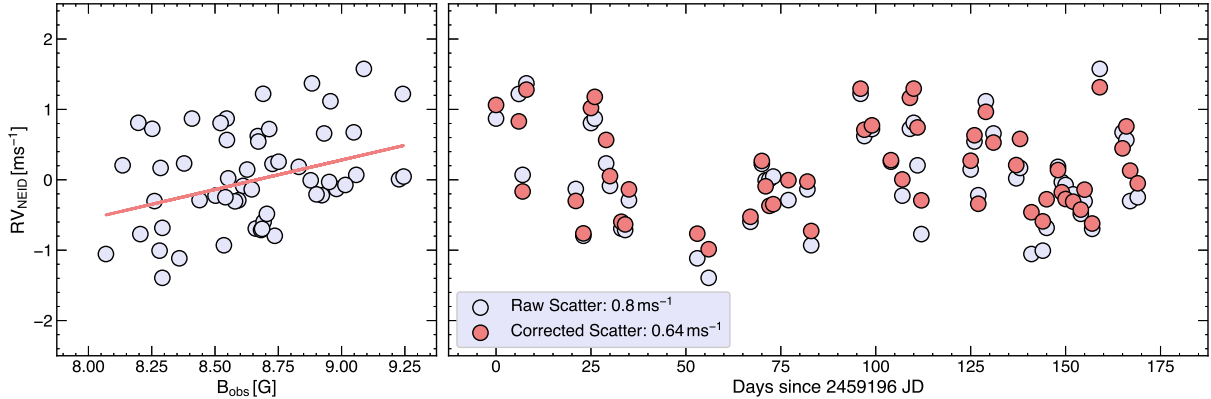}
    \caption{Ground-based RV measurements before and after correction for correlation with unsigned magnetic flux. For these \lq{}good\rq{} weather days, where there was low measurable cloud cover at WIYN, the correlation between $RV_\mathrm{NEID}$ and {\bobs} is 0.43, while the correlation between $\Delta RV_\mathrm{model}$ and $RV_\mathrm{NEID}$ is 0.29, and the average daily binned error for $RV_\mathrm{NEID}$ is $\sim$5 {\cms}. We linearly fit the correlation between {\bobs} and $RV_\mathrm{NEID}$ and subtract this from $RV_\mathrm{NEID}$  to reduce the scatter due to the unsigned magnetic flux. The scatter reduces from $\sim$80 {\cms} to $\sim$65 {\cms} after this correction.}
    \label{fig: Bobs-rv}
\end{figure*}

Performing a linear regression of NEID RVs against measured unsigned magnetic flux (see Figure~\ref{fig: Bobs-rv}), the NEID RV measurement RMS decreases from $\sim$80 {\cms} to $\sim$60 {\cms} over the five month span (Figure \ref{fig: Bobs-rv}) -- an improvement of $\sim$50 {\cms} in a quadrature sum sense. This simplistic activity decorrelation implies there is much promise for using unsigned magnetic flux to reduce RV variability, and we await additional observations to ensure this correlation remains consistent as the Sun enters a more active phase of the magnetic cycle.

\subsection{Filling Factor} \label{sec: f}

The magnetic filling factor, defined as the fraction of the observed solar disk that is active (corrected for foreshortening), is the second metric we studied and compared to the NEID RVs:

\begin{equation}
	f = \frac{\sum_{ij} W_{ij}}{N_\mathrm{pix}}
\end{equation} where $W_{ij}$ is the magnetic weighting array and $W_{ij} = 1$ for active pixels and 0 for pixels identified as inactive.

\begin{figure} [htb!]
    \centering
    \includegraphics[width=\columnwidth]{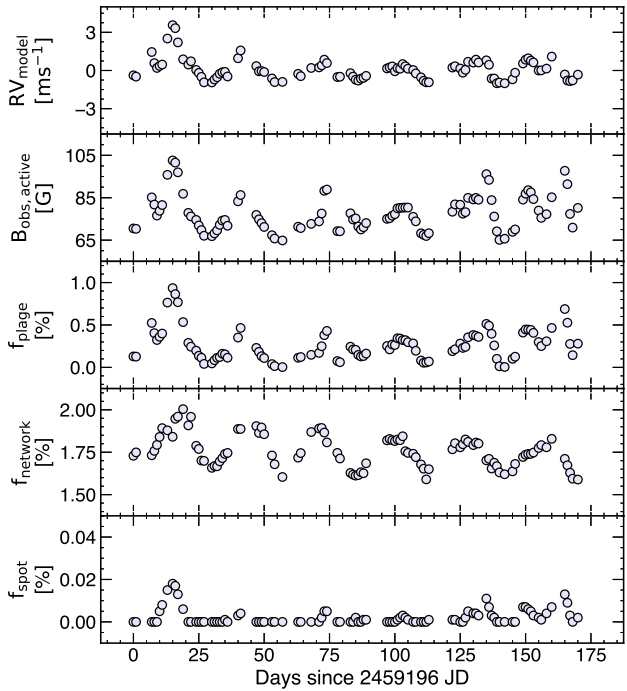}
    \caption{Comparison of faculae in condensed plage areas, faculae in the diffuse network, and sunspot filling factors (\%) with $\Delta RV_\mathrm{model}$ (m s$^{-1}$) and $|\hat{B}_{\mathrm{obs}, \, \mathrm{active}}|$, the unsigned flux due to active regions. We see the strongest correlation between $f_\mathrm{plage}$ and $|\hat{B}_{\mathrm{obs}, \, \mathrm{active}}|$ (Spearman coefficient of 0.97) and note the low percentage of sunspots on the surface as this time span falls during the minimum of Solar Cycle 25. The maximum sunspot percentage on the solar surface in this time span is 0.026\% while faculae cover between 1.5 - 3\% of the solar disk.}
    \label{fig: magnetic}
\end{figure}

The Spearman correlation coefficient between {\bobs} and $f$ is 0.93 and we also find strong correlation between $f$ and $\Delta \hat{v}_\mathrm{conv}$ (0.97), and $\Delta RV_\mathrm{model}$ (0.96) for the full five month time span. We compare the unsigned flux due to active regions with the three filling factors calculated (see Figure~\ref{fig: magnetic}). The dominant feature driving the active region flux is the condensed faculae regions known as plage. Plage are bright regions on the solar surface typically found near sunspots that make up the majority of polarity in solar active regions \citep{Buehler-2019}. MHD simulations show that the amount of thick flux tubes in plage regions is small, while flux tubes in the faculae network expand more quickly than those in plage regions, and that the continuum intensity of bright regions strongly correlates with magnetic field strength \citep{Rohrbein-2011, Danilovic-2013}. 

\section{Comparing \emph{SDO} measurements with NEID CCF metrics}\label{sec: lines}

In addition to comparing our space-based observables to integrated ground-based RVs, we also compared them to RVs calculated using different spectral line masks. These filtered masks were used to try and isolate the most and least activity-sensitive features in the NEID solar spectrum. To test this, we built several physically-motivated line masks to compare the effects of magnetic activity on different groups of spectral lines using our \emph{SDO}-derived solar activity proxies. 

We derived RVs using these tailored masks by calculating cross-correlation functions (CCF) in the same manner as the standard NEID pipeline. These masks were selected based on filtering for line species and line depth, and in all cases our starting line list was the  standard ESPRESSO G2 line mask {\espressofoot}. Once computed, the CCFs were fit with Gaussian functions to determine the averaged spectrum velocity for that particular mask.  

\subsection{Exploration of line depths and the relationship with magnetic observables} \label{sec: line-weights}

The shifts in spectral line profile shape are primarily driven by convection \citep{Dravins-1981} associated with velocity and intensity variations in stellar granules \citep{Hathaway-2000}. As the plasma from active regions (sunspots, faculae, plage) interacts with the solar magnetic field, we see an inhibition of this convective blueshift (see section \ref{sec: vconv}) and this effect varies with line depth \citep{Gray-2009}. Deeper lines are less blue-shifted than shallow lines \citep{Gray-2009} and thus we expect to see a stronger inhibition of the convective blueshift from the mask built with shallow lines. 

We study the effects of deep lines versus shallow lines to determine how these correlate with both the bulk NEID RVs, as calculated by the NEID pipeline using the ESPRESSO G2 mask, and unsigned magnetic flux values computed by {\packagename}. We first applied a binary split between \lq{}deep\rq{} lines and \lq{}shallow\rq{} lines (deep lines having binary mask weights $\geq$ 0.5 and shallow having weights $<$ 0.5) to see if the correlation between the NEID RVs and the unsigned magnetic flux changed measurably (see Figure~\ref{fig: lines}). The RVs calculated using the mask of deep lines show strong correlation with NEID RVs (Spearman correlation coefficient of 0.92), but show weaker correlation with unsigned magnetic flux (correlation coefficient of 0.33). For shallow lines, we similarly found a strong correlation with the NEID RVs (Spearman coefficient of 0.98) and weak correlation with unsigned flux (correlation coefficient of 0.28). The rms ampltiude for the RVs calculated using the depth cut line mask is $\sim$90 {\cms} for the deep lines and $\sim$85 {\cms} for the shallow lines. We also look at the residual scatter between the integrated ground-based RVs (RV$_\mathrm{NEID}$) and the RVs from our curated masks, finding rms scatter of $\sim$22 {\cms} for deep lines and $\sim$10 {\cms} for shallow lines. We find that the RVs calculated using either deep or shallow lines show similar correlations with NEID RVs and unsigned flux, along with similar RMS scatter (Figure~\ref{fig: lines}). 

\subsection{Derivation of ground-based RVs using physically-motivated masks} \label{sec: line-masks}

In addition to studying the variability as a function of line depth, we also constructed masks based on previously published studies that isolate active lines. To attempt to enhance the activity signature in the NEID spectra, we adapted the line list from \citet{Wise-2018} of activity-sensitive lines and recomputed the NEID RVs. This list contains only those lines that showed significant correlations between their line depths and the chromospheric Ca II H\&K index for active stars \citep{Wise-2018}. We then compared these RVs with our \emph{SDO}/HMI pipeline measurements, as well as the \lq{}standard\rq{} NEID pipeline RVs. We found the RVs of these select $\sim$20 lines (those found in the Wise list and ESPRESSO G2 mask) correlate well with the NEID pipeline RVs, and do not show a strong correlation with the \emph{SDO}-derived unsigned magnetic flux. This may be unsurprising, given that \citet{Wise-2018} built this list of activity-sensitive lines through observations of stars that are more active than the Sun's current activity level and show significantly higher rotationally-modulated behavior. As such, the observed relationship between activity and line variability may not hold nor be measurable during this time of low solar activity.

\begin{figure*} [htb!]
    \centering
    \includegraphics[width=\textwidth]{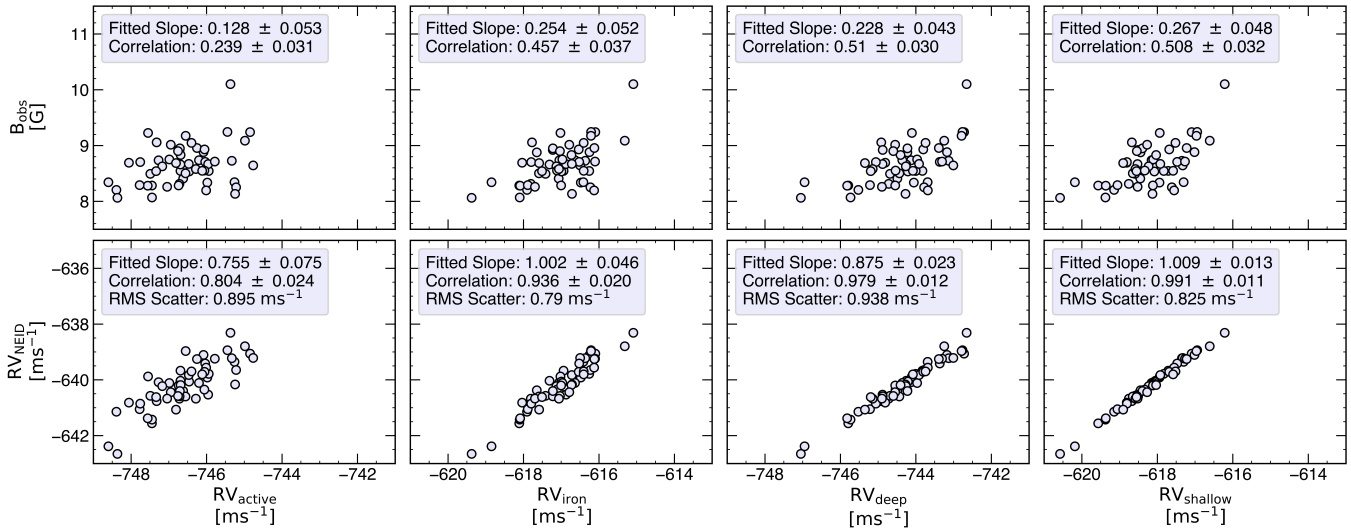}
    \caption{Comparison of NEID RVs derived using different cross-correlation line masks (columns) compared to the \emph{SDO} unsigned magnetic flux (top row) and the bulk NEID RVs (bottom row). Spearman correlation coefficients are shown in the upper left in each panel along with fitted slope values and uncertainties on both measurements. \emph{First column:} Correlations for RVs calculated using activity-sensitive lines from \citet{Wise-2018}. There is moderately strong correlation with the NEID pipeline RVs, but only a weak correlation with unsigned magnetic flux. \emph{Second column:} Correlation for RVs derived using only Fe I lines in the ESPRESSO G2 mask, showing stronger correlation with both the bulk NEID RVs and the unsigned magnetic flux. \emph{Third column:} Correlation for RVs derived using deep lines in the ESPRESSO G2 mask, showing very strong correlation with the bulk RVs. \emph{Fourth column:} Correlation for RVs derived using shallow lines in the ESPRESSO G2 mask, showing the strongest correlation with the bulk RVs. All error bars report only photon noise, excluding any instrument systematics or additional stellar jitter (5-7 {\cms}).}
    \label{fig: lines}
\end{figure*}

Figure \ref{fig: lines} shows the correlation between the RVs derived using different line masks and the \emph{SDO}-derived unsigned magnetic flux. The left panel shows the correlations for the activity-sensitive line list \citep{Wise-2018} while the second column displays the correlations for the mask built using only Fe I lines within the ESPRESSO G2 mask. We filtered specifically for Fe I lines to compare against the SDO-derived RVs, which use a single Fe line to compute the majority of the observables used in {\packagename}. This Fe I line list contains significantly more features than the mask constructed using features from \citet{Wise-2018}, yet we find similar scatter in the RV time series in both masks, implying neither list is significantly more activity sensitive. We do find that the integrated RVs from the Fe I lines show mildly stronger correlation with both the bulk NEID RVs and unsigned magnetic flux in comparison to the activity-sensitive lines from \citet{Wise-2018}. 

\subsection{Comparing measurables from NEID Cross Correlation Functions} \label{sec: ccf-metrics}

Beyond integrated RV measurements, we also studied an assortment of CCF parameters to determine which metrics best trace magnetic field variability. The primary CCF metrics we compared are: fitted amplitude, full width at half maximum (FWHM), skew, and integrated area below the line profile, in addition to calculating the RV shift via Gaussian fit of our CCF. Comparing the variation of these metrics with magnetic observables over time provides insight into the effects driving the CCF shape changes we observe. 

\begin{figure*} [htb!]
    \centering
     \includegraphics[width=\textwidth]{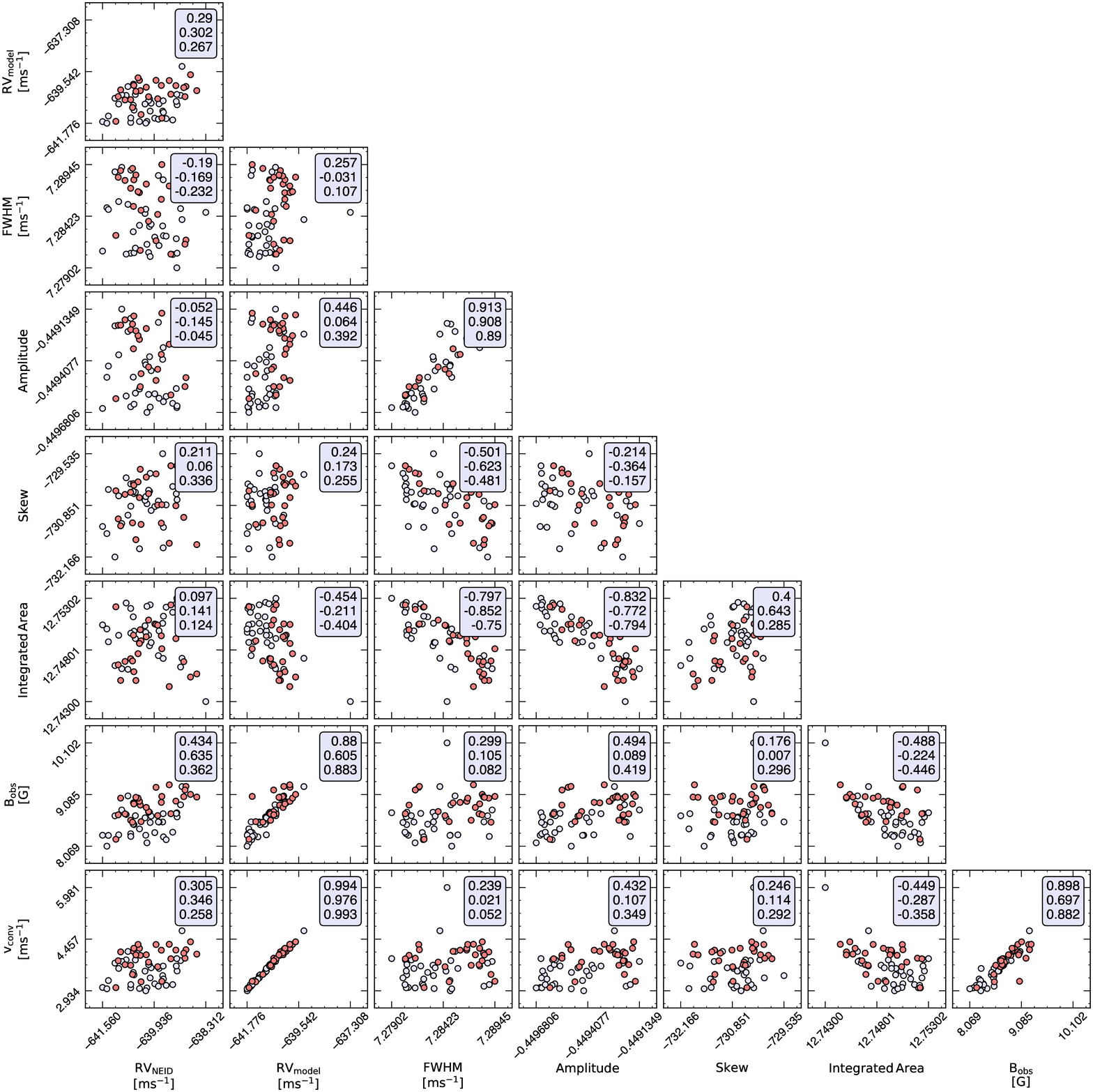}
    \caption{Corner plot showing correlation between space-based magnetic observables and ground-based CCF metrics derived using the full G2 ESPRESSO mask. The time frame for the dataset shown is December 2020 - May 2021. Days of high activity (measurable spots) are shown as dark pink dots and days of quiet-Sun are shown in light lavender. Correlation coefficients are listed in the legends in each sub-plot for the full dataset (top value), high activity days (middle), and quiet-Sun days (bottom), respectively. When comparing the space and ground-based data, we see the strongest correlation between unsigned magnetic flux, {\bobs}, and the CCF amplitude and integrated area measurements.}
    \label{fig: corner_ccfs}
\end{figure*}

Figure~\ref{fig: corner_ccfs} shows the full suite of computed CCF measurements compared to the {\packagename} data products. We looked at CCF metrics using the full G2 ESPRESSO mask which covers wavelengths from 3700 - 7900 {\AA} to calculate RV variations and the list of metrics outlined previously. We find that certain CCF metrics serve as better proxies for magnetic activity due to their strong correlation with unsigned magnetic flux. The SDO data allow us to isolate days with higher magnetic activity, namely days with larger magnetic filling factors, and compare these results to days of \lq{}quiet-Sun\rq{}. 

 We find the strongest proxies for magnetic activity to be integrated area and amplitude, each showing moderate correlation coefficients ($\sim$0.5, see Figure~\ref{fig: corner_ccfs}) when compared to {\bobs}. This broadly supports the results of \citet{Costes-2021}, specifically in regards to the CCF FWHM and integrated area measurements, though we note the base level of solar activity is different between these two studies. The moderately strong negative correlation between integrated area and unsigned magnetic flux is also supported by the work of \citet{CollierCameron-2019} and \citet{Costes-2021}, where integrated area was found to be a strong tracer for the evolution of the magnetic network due to the variation in CCF area showing little to no rotational modulation. The lack of rotational modulation of the integrated area metric implies that its variation is likely driven by axisymmetrically distributed structures over the solar surface, specifically from the circulation of dispersed magnetic flux elements from regions that were once active \citep{CollierCameron-2019}. For days with sunspots, the correlation between photometric velocity and unsigned flux increases while the correlation between convective velocity and unsigned flux decreases. This shows the effect of the rotationally modulated spots on the photometric velocity component as discussed in section \ref{sec: vphot} and supports the correlation between spot factor and jumps in unsigned active magnetic flux as seen in Figure \ref{fig: magnetic}.

\begin{figure*} [htb!]
    \centering
    \includegraphics[width=0.7\textwidth]{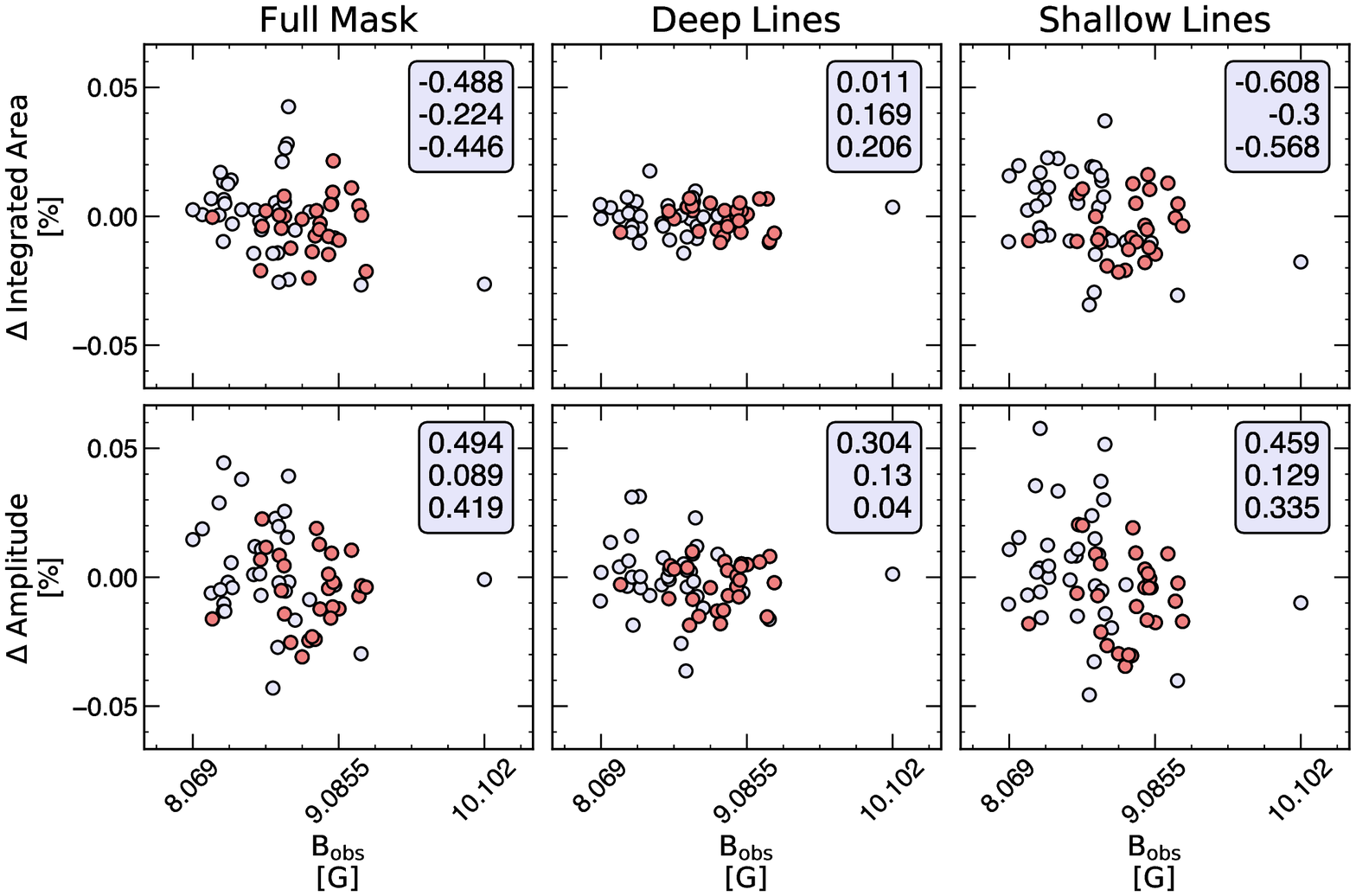}
    \caption{Comparison of the two CCF metrics (amplitude and integrated area) that correlate most strongly with unsigned magnetic flux as a function of line depth (deep having CCF mask weights $\geq$ 0.5, and shallow having weights $<$ 0.5). We show a comparison of the percent change of these metrics against the unsigned magnetic flux. The correlation coefficients are listed in the legend for the full dataset (top value), high activity days (middle), and quiet-Sun days (bottom), respectively. The dark red dots show days of high activity, while the light lavender points are quiet-Sun days.}
    \label{fig: strong_ccf_metrics}
\end{figure*}

To explore the activity behavior as a function of spectral line depth, we created several CCF masks using different depth cuts. As previously described (section~\ref{sec: line-weights}), we use the CCF mask line \lq{}weights\rq{} listed in the ESPRESSO mask as proxies for relative depth. We then recompute the CCF and resultant RVs for all lines within each depth cut. We find that across all depth masks, we see a strong correlation between the derived RVs and bulk NEID pipeline RV measurements (indicating the majority of lines are shifting in a similar manner). For CCF metrics derived with the shallow line mask, we see stronger correlation between unsigned flux and both amplitude and integrated area in comparison to the metrics calculated with the deep line mask as shown in Figure~\ref{fig: strong_ccf_metrics}. In addition, our observed correlations broadly supports previous results from \citet{Meunier-2017}, and \citet{Reiners-2016} of stronger correlation between the CCF RVs from only the shallow lines and the calculated convective velocity component.

\section{Application of {\packagename} to archival planetary transits} \label{sec: transits}

Since the launch of \emph{SDO} in 2010, three planetary solar transits have been imaged by \emph{SDO} instruments -- Mercury in 2016 and 2019, and Venus in 2012. Here we detail our attempts at recovering the corresponding Rossiter–McLaughlin (RM) signals due to the transits of Venus and Mercury. The measurement of these transits serve as proof of concept of using spatially resolved disk images to calculate disk-integrated RVs at precisions currently unattainable from the ground. We apply our \texttt{SolAster} pipeline methods to these time frames in the attempt to 1) recover the effect of the planetary transit on the RV measurements and 2) empirically estimate the precision floor of our constructed \lq{}Sun-as-a-star\rq{} RVs. Beyond testing our pipeline, these RV measurements also showcase the magnitude of the effect that stellar activity has on our ability to detect small RV signals over short timescales. 

For the all three transits, we used \emph{SDO}/HMI data products at a two minute cadence and computed the unsigned magnetic flux, convective and photometric velocities. We then reconstruct the overall model RV variation based on the parameterization outlined in Section~\ref{sec: rv-calc}.


\subsection{Mercury Transits} \label{sec: mercury-2016}

Since the start of the \emph{SDO} observation period, there have been two Mercury transits on May 9, 2016 and on November 11, 2019. Using \texttt{SolAster} we attempt to recover the RM signal of these planetary transits to understand the noise floor (both astrophysical and instrumental) of RV measurements due to the extremely low amplitude expected from these transits.

\subsubsection{Mercury 2016 Transit}
\begin{figure} [htb!]
    \centering
    \includegraphics[width=\columnwidth]{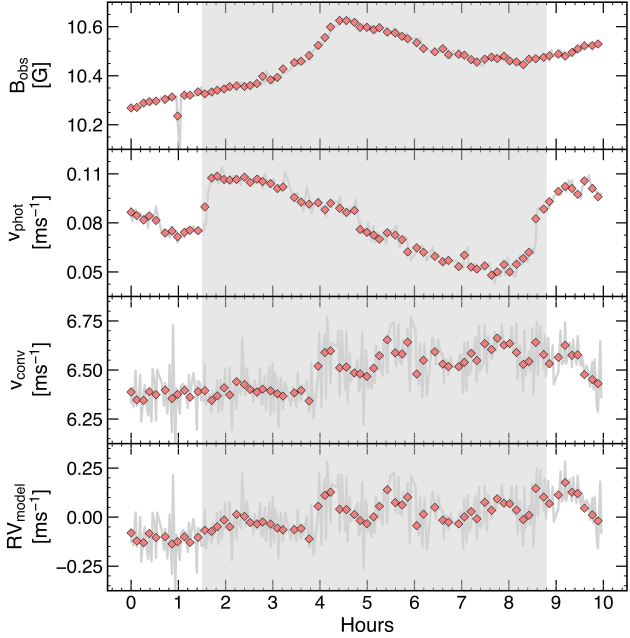}
    \caption{Calculated RVs for the eight hour Mercury transit on May 9, 2016. \emph{Top panel:} Unsigned magnetic flux over the period of the transit. \emph{Second panel:} Photometric velocity from the 2016 transit.  \emph{Third panel:} Convective velocity from the 2016 transit. \emph{Fourth panel:} Model RV variation constructed from \emph{SDO}/HMI data during the planet's transit. For all panels, the red diamonds show the 5.25 minute binned RVs (averaging out p-mode oscillations) while the background grey signal shows the 2 minute cadence of our calculations. Even with the low-level of underlying surface activity, the integrated RV model is dominated by $\Delta \hat{v}_\mathrm{conv}$.}
    \label{fig: mercury-2016}
\end{figure}

The RV signal induced during the Mercury 2016 transit was expected to be on the $\sim$5 {\cms} level, and our constructed RV model from \emph{SDO}/HMI (using the weighting factors in Table~\ref{tab: scale}) shows significantly higher variability. Through observations of the Sun, we have established that solar RV variations are driven by bright faculae (in regions of concentrated plage). From long-term surveys, such as the Mt Wilson HK project \citep{Baliunas-1988}, we know the surfaces of old, slowly rotating Sun-like stars are faculae-dominated \citep{Radick-1983, Lockwood-1984}. These surveys monitored the optical photometric variations and the Ca II H\&K over decades for FGK stars noting that brightness increases as a function of activity, just as observed for the Sun throughout its magnetic cycle. Therefore, these stars are faculae rather than spot dominated and we expect RV variations to be driven by the suppression of the convective blueshift, as seen for the Sun. This is generally consistent with our observed $\Delta \hat{v}_\mathrm{conv}$ in Figure~\ref{fig: mercury-2016}, which has significantly more scatter than$\Delta \hat{v}_\mathrm{phot}$. This result serves as a glaring example of the ability of stellar activity to degrade RV sensitivity to planetary signals, even over short timescales. 


\subsubsection{Mercury 2019 Transit}

\begin{figure} [htb!]
    \centering
    \includegraphics[width=\columnwidth]{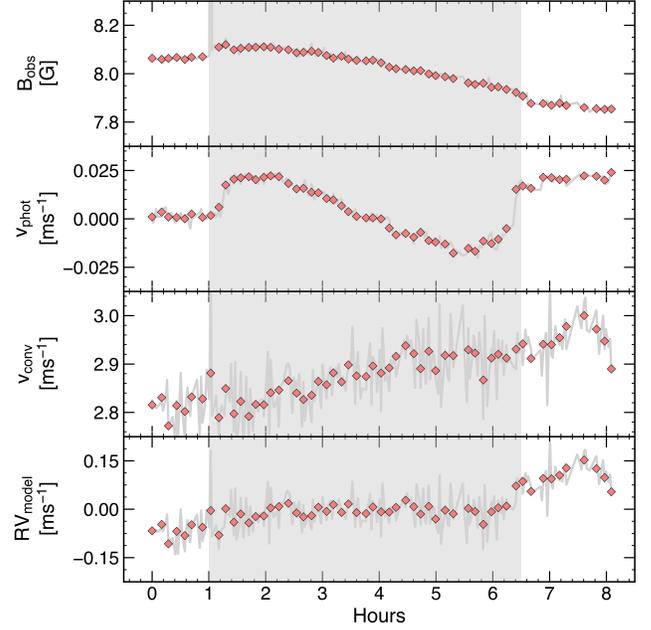}
    \caption{Calculated RVs and magnetic observables for the Mercury transit on November 11, 2019. \emph{Top panel:} Unsigned magnetic flux over the period of the transit showing the low level of activity during this time period. \emph{Second panel:} Photometric velocity component from the Mercury 2019 transit. \emph{Third panel:} Convective velocity component from the Mercury 2019 transit. \emph{Fourth panel:} Model RV variation constructed from \emph{SDO}/HMI data during the planet's transit using the \texttt{SolAster} pipeline as outlined in Section~\ref{sec: rv-calc}. For all panels, the red diamonds show the 5.25 minute binned RVs (averaging out p-mode oscillations) while the background grey signal shows the 2 minute cadence of our calculations.}
    \label{fig: mercury-2019}
\end{figure}

Identical to the construction of the model RVs for the Mercury 2016 transit, we also look at the November 11, 2019 transit of Mercury. This transit period was clear of sunspots and occurred while the Sun moved out of the absolute solar minimum of Solar Cycle 24 which occurred in October 2019, just prior to the transit. These low activity conditions provided an ideal background for the recovery attempt of this $\sim$5 {\cms} signal. We show our recovery attempt along with the component velocities in Figure~\ref{fig: mercury-2019}. This transit occurred at an even lower level of solar activity than the 2016 Mercury transit, but the overall RV variability is still largely dominated by the convective velocity component. 

\subsection{Venus 2012 Transit} \label{sec: venus}

\begin{figure} [htb!]
    \centering
    \includegraphics[width=\columnwidth]{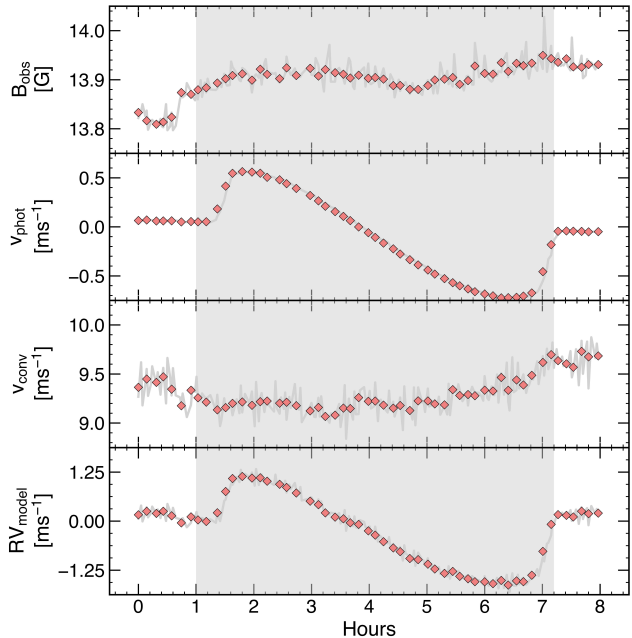}
    \caption{Calculated RVs for the Venus transit from June 5 to June 6, 2012. \emph{Top panel:} Unsigned magnetic flux over the period of the transit. \emph{Second panel:} Photometric velocity component from the Venus 2012 transit. \emph{Third panel:} Convective velocity component from the Venus 2012 transit. \emph{Fourth panel:} Model RV variation constructed from \emph{SDO}/HMI data during the planet's transit using the \texttt{SolAster} pipeline as outlined in Section~\ref{sec: rv-calc}. For all panels, the red diamonds show the 5.25 minute binned RVs (averaging out p-mode oscillations) while the background grey signal shows the 2 minute cadence of our calculations. }
    \label{fig: venus}
\end{figure}

The Venus transit occurred during a period of high activity in the Solar Cycle, however this specific day had relatively low solar activity with a small sunspot and facular filling factor which remained consistent throughout the transit period. The transit occurred from 22:09 UTC on June 5, 2012 until 04:49 UTC on June 6, 2012. Using the \texttt{SolAster} pipeline, we calculate the RV components and reconstruct the model RV variation as outlined in Section~\ref{sec: rv}. We show the overall velocity signal is largely dominated by the RM signal, rather than the convective velocity component (see Figure~\ref{fig: venus}), allowing for a clean recovery of the RM waveform.

\section{Discussion} \label{sec: discussion}

Our Python based, publicly available \emph{SDO}/HMI analysis pipeline (\packagename) allows us to calculate both magnetic observables and \lq{}Sun-as-a-star\rq{} RV variations using space-based data for comparison with ground-based measurements. Moving forward, these data products will aid in studies aimed at deriving new stellar activity indicators explorations in ground-based spectra. By looking at correlations between the space-based data and ground-based measurements from RV facilities such as NEID, we hope to find stronger proxies for stellar activity in Sun-like stars, which would help to improve planet detection sensitivity in future RV surveys. Leveraging the \emph{SDO}/HMI and NEID data, there are a variety of paths we aim to explore in future studies, including:
\begin{itemize}

\item Comparing line-by-line \citep{Dumusque-2018} and integrated CCF metrics with space-based observables, and more closely explore the metrics that show the most promise as activity proxies (CCF integrated area and amplitude being two examples that show promise based on our preliminary study). 

\item Exploring the wavelength dependence of correlations between NEID measurements and \emph{SDO}-pipeline calculations both in RVs and CCFs. This could allow us determine whether the line shape variability is driven by Zeeman effects \citep{Reiners-2013}. If this is the case, we would expect to see a stronger correlation between unsigned magnetic flux and the CCF depth/integrated area in the redder portions of the spectrum. If instead the variation is driven by the inhibition of the convective blueshift, then we expect to see stronger correlation in the blue lines \citep{Reiners-2013}.

\item Detrending NEID RVs against the unsigned magnetic flux ({\bobs}) using more complex parameterizations: including the FF' \citep{Aigrain-2012} technique and Gaussian Processes \citep{Haywood-2014} and others. While our preliminary linear detrending (Figure~\ref{fig: Bobs-rv}) did improve the scatter for a subset of the NEID data, a more robust use of the \emph{SDO} data, looking at the solar rotation period and life time of sunspots, to detrend the RVs could further decrease the activity signal.

\item Revisiting our analysis on times with heightened solar activity. The time span used in this study (December 2020 - May 2021) was during a period of low solar activity. While there were times with sunspots and slightly increased activity levels, the overall activity level was very low, both for the Sun itself and in comparison with other Solar-type stars. Studying periods of higher solar activity (which we are entering now) that have complimentary ground-based spectra (e.g. 2013-2015, with HARPS-N \citep{Dumusque-2020}) would be useful for quantifying how the level of solar activity affects the correlations between the ground-based metrics and space-based observables. This would also allow us to fine tune our scaling factors (Table~\ref{tab: scale}) and derive more precise model RVs. This would also allow us to better understand how the amplitudes and relative contributions of the two velocity components, $\Delta \hat{v}_\mathrm{conv}$ and $\Delta \hat{v}_\mathrm{phot}$, vary as a function of time and activity level.
\end{itemize}

\section{Conclusion} \label{sec: conclusion}

Using \emph{SDO}/HMI data products and ground-based spectra, we studied the \lq{}Sun-as-a-star\rq{} to estimate RV variations due to forms of solar activity. In doing so, we developed a standalone solar data analysis package, {\packagename}, using the methods outlined by \citet{Haywood-2016} and \citet{Milbourne-2019} to compute model RVs, and validated our results against these previously published datasets. We also calculated additional magnetic observables to search for any correlations between these measurements and ground-based RVs from NEID. We found that while the RV variation is driven by the combination of convective blueshift suppression and the rotational Doppler imbalance due to sunspots and plage, the dominant component is due to the convective blueshift suppression, confirming previous results by \citet{Meunier-2010b} and others. We found the RMS scatter of both $\Delta \hat{v}_\mathrm{conv}$ and $\Delta \hat{v}_\mathrm{phot}$ are lower than the results of \citet{Haywood-2016} and \citet{Milbourne-2019} due to our study taking place during a period of minimal solar activity. 

We found that plage regions are the dominant driver of the observed magnetic flux and convective blueshift suppression, as seen in Figure \ref{fig: magnetic}, and the overall RV variation is dominated by the convective blueshift suppression. The photometric contribution, primarily affected by bright active regions and sunspots, is quite minimal and strongly dependent on sunspot filling factor. These conclusions support previous work on this topic by \citet{Meunier-2010b}, \citet{Haywood-2016}, \citet{Milbourne-2019}, and \citet{Haywood-2020}. 

Filtering the NEID data for only days with optimum observing conditions, we find a strong correlation between the ground-based RVs, model RVs, and unsigned magnetic flux. There is a strong correlation between $B_{\mathrm{obs}}$ and the NEID RVs, which we removed via a linear decorrelation to improve the RMS scatter down to the $\sim$60 {\cms} level. 

To better understand which surface features drive active region magnetic flux, we compared a variety of activity indicators and proxies for solar activity. We found that large facular regions known as plage are the dominant component driving the temporal flux variation. Additionally, we found a strong correlation between these magnetic observables and RV variations, aligning with current work in this field \citep{Haywood-2016, Haywood-2020}. This provides additional evidence for the exploration of unsigned magnetic flux as a proxy for stellar activity, especially in Sun-like stars. 

We also investigated correlations between spectral line shape and magnetic activity indicators. We built a variety of physically motivated line masks to better quantify the affect of magnetic activity on different spectral line parameters. We found that RVs calculated using masks built from either deep lines or shallow lines show similar correlation with both ground-based NEID RV pipeline measurements and unsigned magnetic flux in comparison to RVs from shallow lines. 

Using {\packagename}, we are able to recover the planetary RM signal for the Venus 2012 transit at high SNR (see Figure~\ref{fig: venus}). However, the planetary RM signals in both the 2016 and 2019 Mercury transits are dwarfed by activity signals, which dominate the RV noise floor over during transit. While the 2019 Mercury transit took place during a period of low solar activity, we are still unable to measure the RM signal due to colluding noise from solar activity at levels $\sim$2-4$\times$ the RM signal. We used these transit events to both empirically gauge the noise floor of our integrated RV measurements, as both signals are expected to be of low amplitude ($<$ 50 {\cms}), and study activity signals at short ($<$8 hour) timescales, finding that even at low activity levels we were unable to recover the RM signal due to the Mercury transit highlighting the incredibly small amplitude of the signal.

While a simple linear detrending with unsigned flux enabled a measurable improvement in a subset of NEID RVs down to the $\sim$60 {\cms} level, there are still additional avenues of exploration required in order to reach the $\sim$10 {\cms} sensitivity needed to detect Earth-like planets orbiting Sun-like stars. We are able to recover a planetary transit (RM signal) with an amplitude on the 10's of {\cms} level, showing the promise of using this method to improve our understanding of stellar activity and its effect on RV measurements at a variety of timescales. With more ground-based observations from NEID and the upcoming increase in solar activity, we will be able to continue improving our modeling methods and more finely hone our understanding of how specific types of stellar activity affect ground-based high resolution spectra.

\section{Acknowledgements}
The research was carried out at the Jet Propulsion Laboratory, California Institute of Technology, under a contract with the National Aeronautics and Space Administration (80NM0018D0004). This research used version 3.1 \citep{Sunpy} of the SunPy open source software package.

Based on observations at Kitt Peak National Observatory, National Optical Astronomy Observatory, which is operated by the Association of Universities for Research in Astronomy (AURA) under a cooperative agreement with the National Science Foundation. 
These results are based on observations obtained with NEID on the WIYN 3.5m Telescope. WIYN is a joint facility of the University of Wisconsin–Madison, Indiana University, NSF’s NOIRLab, the Pennsylvania State University, Purdue University, University of California, Irvine, and the University of Missouri. The authors are honored to be permitted to conduct astronomical research on Iolkam Du'ag (Kitt Peak), a mountain with particular significance to the Tohono O'odham.

\software{
\texttt{astropy} \citep{AstropyCollaboration2018_astropy},
\texttt{barycorrpy} \citep{Kanodia2018_barycorrpy}, 
\texttt{matplotlib} \citep{Hunter2007_matplotlib},
\texttt{numpy} \citep{vanderWalt2011_numpy},
\texttt{pandas} \citep{McKinney2010_pandas},
\texttt{scipy} \citep{Virtanen2020_scipy},
\texttt{corner} \citep{corner},
\texttt{SunPy} \citep{Sunpy}
}

\bibliographystyle{aasjournal}
\bibliography{ms}{}

\end{document}